\documentclass[twocolumn,tighten,twocolappendix]{aastex631}
\usepackage{gensymb}
\usepackage{amsmath}
\usepackage{booktabs}
\usepackage{graphicx}
\usepackage{natbib} 
\usepackage{amssymb}
\usepackage{bm}

\hypersetup{linkcolor=magenta,citecolor=blue,filecolor=blue,urlcolor=blue}

\defcitealias{f16}{F16}
\defcitealias{mar23}{Mar23}
\defcitealias{gh21}{GH21}
\defcitealias{mal12}{Mal12}
\defcitealias{r13}{R13}
\defcitealias{ss18}{SS18}
\defcitealias{se19}{Se19}
\defcitealias{w20}{W20}
\defcitealias{z23}{Z23}
\defcitealias{Dong2023}{Paper I}

\newcommand{\gaia}{\textit{Gaia}}

\newcommand{\vlsr}{$v\rm_{\scriptscriptstyle LSR}$}
\newcommand{\kms}{$\rm km\,s^{-1}$}
\newcommand{\masyr}{$\rm mas\,yr^{-1}$}
\newcommand{\pmra}{$\mu_{\alpha^{*}}$}
\newcommand{\pmdec}{$\mu_{\delta}$}
\newcommand{\plx}{$\varpi$}
\newcommand{\g}{$G$}

\newcommand{\msun}{$M_\sun$}
\newcommand{\cli}{Class \uppercase\expandafter{\romannumeral1}}
\newcommand{\clii}{Class \uppercase\expandafter{\romannumeral2}}
\newcommand{\cliii}{Class \uppercase\expandafter{\romannumeral3}}
\newcommand{\clitoii}{Class \uppercase\expandafter{\romannumeral1}/\uppercase\expandafter{\romannumeral2}}
\newcommand{\cliitoiii}{Class \uppercase\expandafter{\romannumeral2}/\uppercase\expandafter{\romannumeral3}}
\newcommand{\uco}{$^{12}$CO}
\newcommand{\lco}{$^{13}$CO}
\newcommand{\ltco}{C$^{18}$O}
\newcommand{\multilines}{$^{12}$CO, $^{13}$CO, and C$^{18}$O}
\newcommand{\ra}{$\alpha$}
\newcommand{\dec}{$\delta$}
\newcommand{\eplx}{$ \sigma_\varpi $}
\newcommand{\epmra}{$ \sigma_{\mu_{\alpha^{*}}} $}
\newcommand{\epmdec}{$ \sigma_{\mu_{\delta}} $}

\shorttitle{3D Morphology and Motions of the CMa}
\shortauthors{Dong et al.}

\graphicspath{{./}}

\begin{document}

\title{3D Morphology and Motions of the Canis Major Region from \gaia~DR3}
\correspondingauthor{Ye Xu}
\email{xuye@pmo.ac.cn}

\author{Yiwei Dong}
\affiliation{Purple Mountain Observatory, Chinese Academy of Sciences, Nanjing 210023, People's Republic of China}
\affiliation{School of Astronomy and Space Science, University of Science and Technology of China, Hefei 230026, People's Republic of China}
\author[0000-0001-5602-3306]{Ye Xu}
\affiliation{Purple Mountain Observatory, Chinese Academy of Sciences, Nanjing 210023, People's Republic of China}
\affiliation{School of Astronomy and Space Science, University of Science and Technology of China, Hefei 230026, People's Republic of China}
\author{Chaojie Hao}
\affiliation{Purple Mountain Observatory, Chinese Academy of Sciences, Nanjing 210023, People's Republic of China}
\author[0000-0001-7526-0120]{Yingjie Li}
\affiliation{Purple Mountain Observatory, Chinese Academy of Sciences, Nanjing 210023, People's Republic of China}
\author{Dejian Liu}
\affiliation{Purple Mountain Observatory, Chinese Academy of Sciences, Nanjing 210023, People's Republic of China}
\affiliation{School of Astronomy and Space Science, University of Science and Technology of China, Hefei 230026, People's Republic of China}
\author[0000-0002-3904-1622]{Yan Sun}
\affiliation{Purple Mountain Observatory, Chinese Academy of Sciences, Nanjing 210023, People's Republic of China}
\author{Zehao Lin}
\affiliation{Purple Mountain Observatory, Chinese Academy of Sciences, Nanjing 210023, People's Republic of China}

\begin{abstract}
The Canis Major (CMa) region {is known for its} prominent arc-shaped morphology, visible at multiple wavelengths.
This study integrates molecular gas data 
with high-precision astrometric parameters of young stellar objects (YSOs) from \gaia~DR3 to {provide the first three-dimensional (3D) insights into the dynamical evolution and star formation history of the CMa region}.
By utilizing the average distances and proper motions of the YSOs as proxies for those of the molecular clouds (MCs), 
we {confirm the presence of} a slowly expanding shell-like morphology 
in the CMa region, with {the estimated} radius of {$47\pm11$} pc and expansion velocity of {$1.6\pm0.7$} \kms.
Further, the dynamical evolution of the shell {supports} its expansion, 
with an expansion timescale of $\sim$4.4 Myr {obtained by} the traceback analysis assuming constant velocities.
Finally, a momentum estimate suggests that at least 2 supernova explosions {(SNe)} are needed to power the observed expanding shell, {reinforcing the previous hypothesis of multiple SNe events}. 
This study effectively combines the CO data with the astrometric data of YSOs from \gaia, 
offering significant support for the future studies on the 3D morphology and kinematics of MCs.
\end{abstract}
\keywords{Star forming regions (1565); Young stellar objects (1834); Molecular clouds (1072); Stellar feedback (1602)}

\section{Introduction} \label{sec:intro}
During their lifetimes, stars leave indelible imprints on the surrounding interstellar medium through processes such as stellar feedback, which is essential for shaping gas structures and driving gas motions \citep[e.g.,][]{krumholz2014,herrington2023}.
Young stellar objects (YSOs) are newborn stars that typically remain closely associated with their natal molecular clouds (MCs) \citep[e.g.,][]{lada1987,gupta2022}.
Recently, it has become a new research trend to combine YSOs to unveil the three-dimensional (3D) morphology and kinematics of MCs, thus helping to decipher the local star formation history \citep[e.g.,][]{orion2018,orion2021}.

This study focuses on the Canis Major (CMa) star-forming region, situated at the Galactic longitude of $ \sim$225\degree, Galactic latitude of $ \sim$$-$1.5\degree, and a distance of $ \sim $1 kpc from the Sun. 
The star formation activity in this region mainly occurs within the CMa OB1/R1 associations that include nearly a hundred OB-type stars {\citep{gh2008}}, over a dozen known open clusters \citep[e.g.,][]{hunt2023}, and 5 \ion{H}{2} regions \citep[S292, 293, 295, S296, and S297;][]{Sharpless_1959,gaylard1984}. 
According to previous CO observations, the CMa region contains abundant molecular material, with the \uco-traced MCs reaching a total size of $ \sim$100 pc and mass of $ \sim$$10^5 $ \msun~(e.g., \citealp{Dong2023}, hereafter, Paper I).
Notably, there are scarcely any overlapping foreground or background star-forming regions towards the CMa, making it a highly desirable target for the detailed studies of 3D morphology and motions.

Based on the two-dimensional (2D) observations, previous studies have conducted in-depth research on the CMa region at different wavelengths. 
Observations of both neutral hydrogen \citep{Herbst1977} and ionized gas \citep{reynolds1978} revealed large-scale expanding shells in CMa.
Subsequently, an organized expansion pattern of OB stars in the sky plane was also discovered \citep{comeron1998}. 
Using multi-wavelength (optical, infrared, radio) images, \citet{fernandes2019} proposed that the most prominent arc-shaped nebula in CMa, Sh 2-296, is part of a large shell with a diameter of $\sim$60 pc, designated the ``CMa shell''.

Compared to the stellar wind and \ion{H}{2} region models, 
{the hypothesis that} supernova explosions (SNe) have produced the shell-like structure and triggered the star formation activity in CMa {is supported by the following evidence.} 
{First}, the few massive stars in CMa are insufficient to ionize the shell \citep{Herbst1977,fernandes2019}.
{Second}, three runaway stars are identified, which may be ejected after the explosions of their massive companions \citep{Herbst1977,comeron1998,fernandes2019}. 
{Third}, the age of YSOs seems to be consistent with the age of supernova remnant (SNR) \citep{gh21}. 
\citet{fernandes2019} traced the past trajectories of three runaway stars and found their mutual origin in proximity to the center of the CMa shell.
Thus, they inferred that the region experiences three successive SNe events $\sim$6, 2, and 1 million years ago, and the CMa shell is shaped by the nested SNRs. 
\citet{ss2021}, in their search for stellar groups in \gaia~Data Release 2 (DR2, \citealp{gaiadr2}) in the CMa region, identified a new open cluster candidate near the shell's center that may be the progenitor cluster of the three runaway stars, further supporting the scenario proposed by \citet{fernandes2019}.

The study of molecular gas morphology and kinematics plays a pivotal role in understanding the dynamical evolution of MCs and star formation history. 
Although 2D images at multiple wavelengths have provided some insights into the shell-like structure of the CMa region, its 3D morphology and motions still remain untangled. 
To this end, we intend to investigate the 3D structures and motions of MCs in the CMa region in conjunction with YSOs. 
It is well established in the literature that YSOs, especially those in \clii~or earlier evolutionary stages, not only remain spatially associated with the natal clouds \citep[e.g.,][]{gutermuth2011,orion2018,gh21,gupta2022}, but also inherit their motions \citep[e.g.,][]{hacar2016,dario2017,orion2021}. 
Therefore, YSOs are reliable tracers of the morphology and kinematics of MCs. 

Extensive searches for YSOs towards the CMa region have been conducted using the advanced infrared data \citep[e.g.,][]{mal12,f16,se19}.
Thus, the collection of archival catalogs yields an adequate sample of YSOs. 
The positional and kinematic information of YSOs is then obtained from the high-precision astrometric data from \gaia~Data Release 3 (DR3, \citealp{gaiadr3}).
For MCs, the Milky Way Imaging Scroll Painting (MWISP) project \citep{Su_2019} delivers high-sensitivity and high-resolution \multilines~(1--0) data that help to resolve the faint or subtle molecular structures in detail. 
Integrating the high-precision astrometric data of YSOs from \gaia~DR3 with high-quality molecular gas data from the MWISP project, we revisit the CMa region from a 3D perspective, shedding light on the origination and star formation history of the region.

This paper is organized as follows.
Section \ref{sec:data} summarizes the molecular gas data and the process of gathering and selecting YSOs. Section \ref{sec:results} presents the 3D shell-like morphology and overall expansion of the CMa region. In Section \ref{sec:discussion}, we discuss the probable origination of the shell through the radial distribution of gas column density and traceback analysis. Finally, a summary can be found in Section \ref{sec:conclusions}.

\section{Data} \label{sec:data}
\subsection{Molecular Clouds} \label{sec:co_data}
\begin{figure*}[ht!]
	\centering
	\gridline{\fig{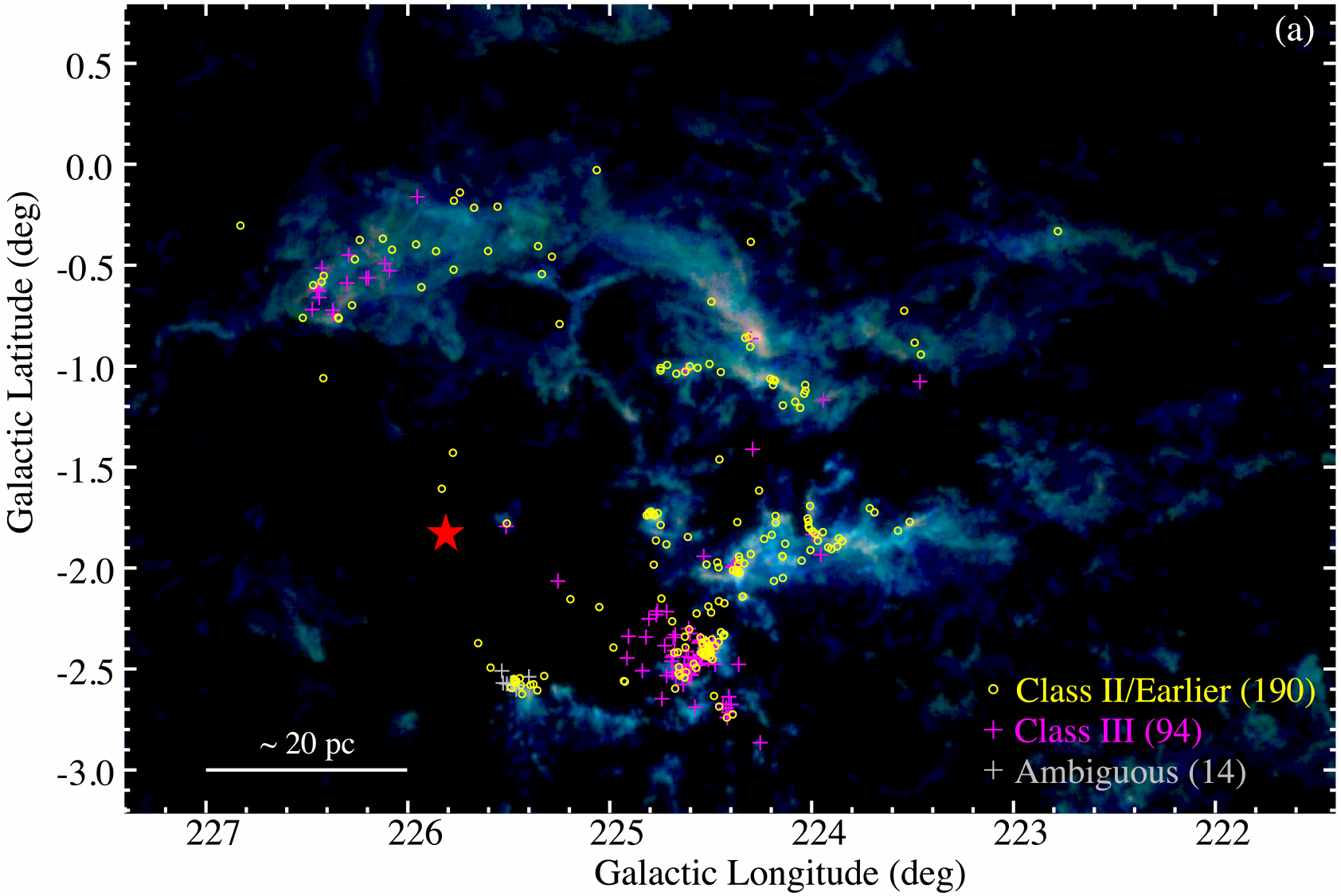}{0.78\textwidth}{}
	\fig{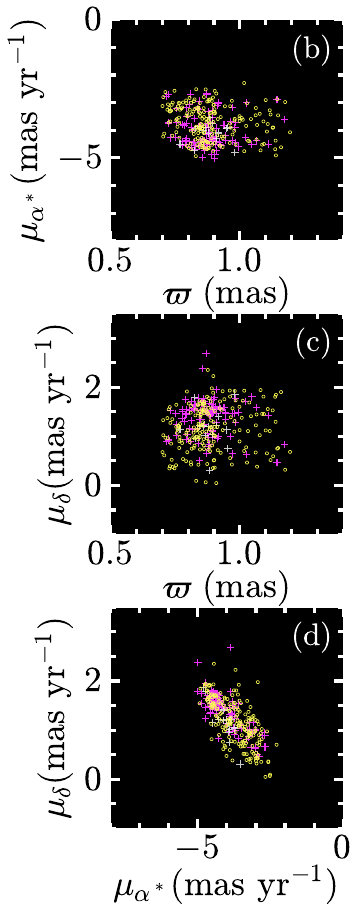}{0.209\textwidth}{}}
	\vspace{-0.8cm}
	\caption{Distributions of YSOs {satisfying all the selection criteria}. (a): the $l$--$b$ distribution of YSOs overlaid on the \uco~(blue), \lco~(green), and \ltco~(red) intensity maps of the three MCs. 
	The yellow circles, magenta crosses, and {gray crosses indicate YSOs in the Class II or earlier classes, Class III, and ambiguous evolutionary stages, respectively}. 
	The numbers of sources are also annotated. 
	The red star represents the center of the CMa shell proposed by \citet{fernandes2019}. 
	(b)--(d): the \plx--\pmra, \plx--\pmdec, and \pmra--\pmdec~distributions of YSOs. 
	\label{fig:m0}}
\end{figure*}

CO has been the most commonly used molecular gas tracer since its first detection \citep{Wilson_1994}.
This study utilizes the \multilines~(1--0) data from the MWISP project \citepalias{Dong2023}, observed with the Purple Mountain Observatory (PMO) 13.7 m millimeter-wave telescope with a beam size of $\sim$50\arcsec~at 115 GHz. 
The median rms noise values are $\sim$0.5 K for \uco~at a velocity resolution of $\sim$0.16 \kms~and $\sim$0.3 K for \lco~and~\ltco~at a velocity resolution of $\sim$0.17 \kms.
The RGB composite image of the molecular gas (blue: \uco, green: \lco, and red: \ltco ) in the CMa region is shown in Figure \ref{fig:m0}, {tracing the previously revealed large-scale} shell structure. 
According to the MC catalog from \citetalias{Dong2023}, the shell is mainly composed of three \uco-traced MCs, MWISP G224.440$-$1.069, MWISP G225.228$-$2.737, and MWISP G224.521$-$2.560, each of which contains numerous relatively dense \lco~and \ltco~structures.
The spatial coverages, LSR velocity ranges, and masses of the three MCs are listed in Table \ref{tab:MCs}.

\subsection{YSOs Sample} \label{sec:yso_data}
\subsubsection{Sample Collection}

\begin{deluxetable*}{cchcccccccch}
	\tablecaption{Three MCs in the CMa Region
	\label{tab:MCs}}
	\tablehead{
	\colhead{{Index}} &\colhead{Name} & \nocolhead{Alias}&\colhead{$ l\rm_{min} $}&\colhead{$ l\rm_{max} $}&\colhead{$ b\rm_{min} $}&\colhead{$ b\rm_{max} $}&\colhead{ $v\rm_{\scriptscriptstyle LSR, min}$}&\colhead{ $v\rm_{\scriptscriptstyle LSR, max}$}&\colhead{Mass$\rm^a$}&\colhead{{$d\rm_{Kim}$$\rm^b$}}&\nocolhead{{Ref.$\rm^c$}}\\
	\colhead{} &\colhead{} &\nocolhead{} &  \colhead{($\degree$)}&\colhead{($\degree$)}&\colhead{($\degree$)}&\colhead{($\degree$)}& \colhead{($\rm km\,s^{-1}$)} &
	\colhead{($\rm km\,s^{-1}$)}&\colhead{(\msun)}&\colhead{(pc)}&\nocolhead{}
	}
	\decimalcolnumbers
	\startdata
	1&	MWISP G224.440$-$1.069 &  North cloud &  221.5 &  227.3 & $-$2.4 &  0.7 &  8.3 &   25.2 &  $ 1.0\times10^5 $ & 1150 &(1--7)\\
	2&	MWISP G225.228$-$2.737 &  South cloud 1&  224.8 &  225.6 & $-$3.0 &   $-$2.5 & 7.3 &  15.9 & $ 3.3\times10^3 $ & 1150&(1)(4)(7)\\
	3&	MWISP G224.521$-$2.560 &  South cloud 2&   224.2 &  224.9 & $-$2.8 &  $-$2.3 &  6.5 &  18.4 & $ 3.8\times10^3 $& 1150&(1)(4)(7) \\
	\enddata
	\tablenotetext{a}{Refers to the molecular mass traced by \uco~(see \citetalias{Dong2023}).}
	\tablenotetext{b}{{Cloud distances from \citet{Kim_2004}}.}
\end{deluxetable*}

Given the comprehensive YSO surveys in the CMa region, we compile the published catalogs of YSOs/YSO candidates (henceforth YSOs, for simplicity).
We maximize the initial sample size with evolutionary stage classifications, while avoiding duplicate catalogs from the same original literature. 
A total of 1,895 YSOs records from 9 references have been included. 
The abbreviations of these references, the covered MCs, and the numbers of employed YSOs are listed in Table \ref{tab:samples}. 
Among them, \citetalias{w20}, \citetalias{z23}, and \citetalias{mar23} cover {numerous star-forming regions in} the Milky Way and the other 6 studies are mainly dedicated to the CMa region.

Based on the first 34 months of observations made by the European Space Agency’s \gaia~mission \citep{gaia2016}, \gaia~DR3 \citep{gaiadr3} provides unprecedented high-precision astrometric data for about 1.5 billion sources, including celestial positions, parallaxes (\plx), proper motions (\pmra~and~\pmdec), and radial velocities for over 33 million stars. 
We cross-match the YSOs sample with \gaia~DR3 in the \gaia~database\footnote{\url{https://gea.esac.esa.int/archive/}} to obtain high-precision astrometric measurements for YSOs. The procedure is as follows.

\begin{enumerate}
	\setlength{\itemsep}{0pt}
	\setlength{\parsep}{0pt}
	\setlength{\parskip}{0pt}
		\item 
		For YSOs with \gaia~EDR3/DR3 IDs from \citetalias{gh21}, \citetalias{z23}, and \citetalias{mar23}, cross-matching is performed directly using \gaia~\texttt{source\_id}.
		\item 
		For YSOs with AllWISE ID from \citetalias{f16} and \citetalias{se19}, as well as YSOs with 2MASS ID from \citetalias{w20} and \citetalias{ss18}, we utilize the \emph{allwise\_best\_neighbour} and \emph{tmass\_psc\_xsc\_best\_neighbour} tables to find the matching \gaia~sources, respectively. 
		\item 
		For YSOs without available IDs from \citetalias{mal12}, \citetalias{r13}, and \citetalias{se19}, cross-matching is conducted using the right ascension (\ra) and declination (\dec) within a radius of 1\arcsec. 
\end{enumerate}

\begin{deluxetable*}{lllccc}
	\tablecaption{Summary of YSOs from the Literature \label{tab:samples}}
	\tablehead{\colhead{Ref.}	&\colhead{{Source Designation}}	&\colhead{{Coverage}}&\colhead{Cloud}&\colhead{$N\rm_{YSO} $} & \colhead{$N_{Gaia} $}
	}
	\decimalcolnumbers
	\startdata
	\citetalias{mal12}	&$\   $--& $l\sim225.4\degree,\,b\sim-2.6\degree $($\sim$$ 7.5\arcmin \times 7.5\arcmin$)&  2
	 &97		&34\\
	\citetalias{r13}	&$\   $--& $l\sim224.5\degree,\,b\sim-2.4\degree $~($\sim$$ 7\arcmin \times 6\arcmin$)&  3
	 &42		&{35}\\
	\citetalias{f16}	&AllWISE&$219\degree\lesssim l\lesssim 231\degree,\,-7\degree\lesssim b\lesssim3\degree $& 1--3 &305	&{160}\\
	\citetalias{ss18}	&2MASS& $l\sim224.7\degree,\, b\sim-2.5\degree $~($\sim$$1.5\degree \times1\degree$)&  3
	 &156	&{146}\\
	\citetalias{se19}	&AllWISE$ ^* $, GLIMPSE360, ...
	& $223.2\degree\lesssim l\lesssim225.8\degree,\,-1.4\degree\lesssim b\lesssim0\degree $&  1 
	&293	&106\\
	\citetalias{w20}	&2MASS, GLIMPSE360& $65\degree\lesssim l\lesssim265\degree,\,-3\degree\lesssim b\lesssim3\degree $
	&  1&751	&{510}\\	
	\citetalias{gh21}	&\gaia~EDR3, 2MASS& $ l\sim226.3\degree,\,b\sim-0.5\degree $~($\sim$$ 40\arcmin \times 20\arcmin$)& 
	1
	&40		&40\\
	\citetalias{z23}	&\gaia~DR3, AllWISE& All-sky  &1--3 &120	&120\\	
	\citetalias{mar23}	&\gaia~DR3& 
	Nearby star-forming regions&  3
	&91		&90\\
	\enddata
	\tablecomments{Column 1: abbreviations of the 9 references, listed from top to bottom as \citet{mal12}, \citet{r13}, \citet{f16}, \citet{ss18}, \citet{se19}, \citet{w20}, \citet{gh21}, \citet{z23}, and \citet{mar23}.
	Column 2: infrared and/or \gaia~IDs given by the catalogs.
	Column 3: the covered {Galactic area}.
	{Column 4: the corresponding MCs}, where {Clouds} 1--3 refer to the MCs {indexes} in the first column of Table \ref{tab:MCs}.
	Column 5: the numbers of YSOs towards the CMa region.
	Column 6: the numbers of YSOs after cross-matching with \gaia~DR3.}
	\tablenotetext{*}{Only a portion of the samples have this ID.}
\end{deluxetable*}

In the first step, one \citetalias{mar23} source with \gaia~DR3 \texttt{source\_id}$=$0 is eliminated from the cross-matching process. 
In the second and third steps, 47 entries are closely inspected, each matched to two \gaia~sources. 
{After considering the magnitude, color, and astrometric parameters from the current \gaia~DR3 data, we can resolve the above confusion for 4 entries}. 
Altogether, these steps yield {1,241} cross-matching entries.
The number of matched samples is shown in Column 6 of Table \ref{tab:samples}.

\subsubsection{Sample Cleanup}
First, following \gaia~technical recommendations\footnote{See details in the technical note GAIA-C3-TN-LU-LL-124-01.}, we apply a cut of $\texttt{ruwe}\leq1.4$ to avoid sources with low-quality astrometric solution (due to binaries, for example), retaining {993} entries. 
Next, we address the duplicate cases. 
Duplicate entries can occur when the same YSO is detected by various IR surveys that may give different source IDs and slightly different coordinates.  
{After removing the duplicate entries using \gaia~\texttt{source\_id}, we get {743} objects.}

{We obtain the classifications for the above {743} objects from the literature.
Among them, {200} have been classified by multiple references, and they are manually checked. 
If the majority of the references provide an identical classification for an object, we keep the object by adopting that classification. 
Otherwise, the object is removed. 
At this stage, there are {722} YSOs in the sample.}

\subsubsection{Sample Selection} \label{sec:select_YSOs}
To ensure more accurate parallax measurements, we apply the criterion of $\texttt{parallax\_over\_error}>3$ by referring to \citetalias{gh21} and obtain {490} sources. 
Initially, {\citet{Kim_2004} determined the distance of \lco~clouds in CMa to be 1150 pc, based on the photometric distance to the CMa OB1 association \citep{Claria_1974}}. 
Recently, high-precision distance measurements to MCs in the CMa region span from $\sim$1057 to 1209 pc \citep[e.g., using extinction method or parallax measurements of YSOs;][]{zucker2019,Dharmawardena2023,z23}. 
To retain the possibly associated YSOs for the CMa clouds, we set a parallax selection criterion of 0.7--1.2 mas {and obtain {319} YSOs. 
Then, we eliminate 21 YSOs that significantly deviate from the spatial ($l$, $b$) distributions of MCs and the overall proper motions of YSOs by visual inspection. 
Figure \ref{fig:m0} illustrates the spatial, parallax, and proper motion distributions of the {298} YSOs that meet all the above criteria}.
%

\begin{deluxetable*}{ccccccccc}
	\tablecaption{\gaia~DR3 parameters of YSOs	\label{tab:catalog}}
	\tablehead{\colhead{\gaia~DR3}	&\colhead{$ l $}&\colhead{$ b $}&\colhead{\plx} & \colhead{\eplx}&\colhead{\pmra}&\colhead{\epmra}&\colhead{\pmdec}&\colhead{\epmdec}\\
	\colhead{\texttt{source\_id}}	&\colhead{(\degree)}&\colhead{(\degree)}&\colhead{(mas)} & \colhead{(mas)}&\colhead{(\masyr)}&\colhead{(\masyr)}&\colhead{(\masyr)}&\colhead{(\masyr)}}
	\decimalcolnumbers	
	\startdata
	3046417902171155328&224.36&$-$2.02&1.01&0.25&$-$3.56&0.21&1.38&0.20\\
	3045239083971370112&226.08&$-$0.42&0.77&0.17&$-$3.21&0.16&0.91&0.15\\
	3046032072370037248&224.51&$-$2.42&1.04&0.17&$-$4.40&0.17&1.69&0.18\\
	3045837772351358336&224.66&$-$2.49&0.87&0.04&$-$4.41&0.04&1.53&0.05\\
	3046027536887737984&224.53&$-$2.43&0.84&0.01&$-$4.41&0.02&1.71&0.02\\
	···&   &  &  &  &  &  &  &  \\
	\enddata
	\tablecomments{This table is available in its entirety in machine-readable form.}	
\end{deluxetable*}

The sample includes {190} sources in \clii~or earlier classes, {94} sources in \cliii, and {14} sources with ambiguous classifications. 
YSOs in \clii~or earlier classes are typically still embedded in circumstellar materials (envelopes and/or disks), while the \cliii~objects represent pre-main-sequence stars with largely diminished circumstellar disks \citep[e.g.,][]{evans2009}. 
Previous studies demonstrate that the former objects are generally good probes of the distances and proper motions of the parental MCs \citep[e.g.,][]{orion2018,orion2021,gupta2022,liudj2024}. 
As shown in Figure \ref{fig:m0} (a), the positions of YSOs in \clii~or earlier evolutionary stages (the yellow circles) coincide well with the emission peaks of the CO gas, and only few YSOs present in the central cavity where molecular gas is scarce. 
{For Class III YSOs, we find that almost all of them are not well spatially associated with MCs and are concentrated in a limited area in the ($l$, $b$) distribution (see Figure \ref{fig:m0} (a)).  
To sum up, we choose 188 YSOs in \clii~or earlier classes to investigate the 3D dynamics of MCs in the CMa region.}

The selected YSOs sample consists of {6} \cli~and {157} \clii~sources, 
2 \emph{Flat} sources\footnote{Refer to YSOs with the slopes of spectral energy distribution between \cli~and \clii, i.e., objects whose emissions arise from both disks and envelopes.} from \citetalias{r13}, 
{23 Class I/II sources} independently offered by \citetalias{z23}\footnote{According to the descriptions of \citetalias{z23}, these sources are Class I/II candidates that meet our needs.}, and
2 \emph{CTT*} sources\footnote{According to \citetalias{mar23}, \emph{CTT*} sources refer to classical T Tauri objects, i.e., YSOs of solar-like spectral types with accretion disks and emission line spectra features.} from \citetalias{mar23}. 
Their \gaia~DR3 parameters are listed in Table \ref{tab:catalog}.
For the sample, the median uncertainties of \plx, \pmra, and \pmdec~are 0.10 mas, 0.10 \masyr, and 0.10 \masyr, respectively.

Figure \ref{fig:m0}~(d), the proper motion space, illustrates that the {distribution} of the selected YSOs (the yellow circles) appears to be congregated around {$\mu_{\alpha^{*}}\sim-4{\rm~mas\,yr^{-1}}$} and $\mu_{\delta}\sim 1\rm~mas\,yr^{-1}$. 
{The overall proper motion feature of the selected YSOs is similar to that of the youngest stellar groups identified by \citet[][see their Figure 7]{ss2021}. 
They reported that the proper motions ($\mu_{\alpha^{*}}, \mu_{\delta}$) of the youngest stellar groups in CMa are roughly in the range of [$-3$, $-5$] \masyr~and [0, 2] \masyr, respectively.}

\section{Results} \label{sec:results}
\subsection{Division of Subregions}\label{sec:subregions}
{Previous CO line surveys divided molecular gas based on their distributions. 
For instance, 
\citet{Kim_2004} conducted an extensive \lco~(1--0) survey covering the CMa region with Nagoya 4 m telescope (beam size $\theta\rm_{MB}\sim2.7\arcmin$ and rms noise level $\sigma\sim0.5$ K). 
They identified about ten \lco~clouds, defined as isolated structures above 5 $\sigma$ contour level.
\citet{Benedettini_2020,Benedettini_2021} used the 12 m antenna of the Arizona Radio Observatory to map \uco~and \lco~(1--0) lines ($\theta\rm_{MB}\sim55\arcsec$ and $\sigma\sim0.3$--1.3 K), covering most of Cloud 1. 
They decomposed it into several cloud structures using the SCIMES algorithm \citep{Colombo2015}, which effectively splits moderate-size consecutive structures in the position-position-velocity (PPV) space. 
\citet{Lin_2021} analyzed MWISP \uco, \lco, and \ltco~data (the same as this study, $\theta\rm_{MB}\sim50\arcsec$ and $\sigma\sim0.3$--0.5 K) for Cloud 1, preliminarily segmenting it into three subregions based on the morphology of \lco~emission. 
In \citetalias{Dong2023}, Clouds 1--3 were identified from the MWISP \uco~data with the DBSCAN algorithm \citep{Yan_2020}, which is designed to detect independent consecutive structures in the PPV space.}

\begin{figure*}[ht!]
	\centering
	\includegraphics[width=1\textwidth]{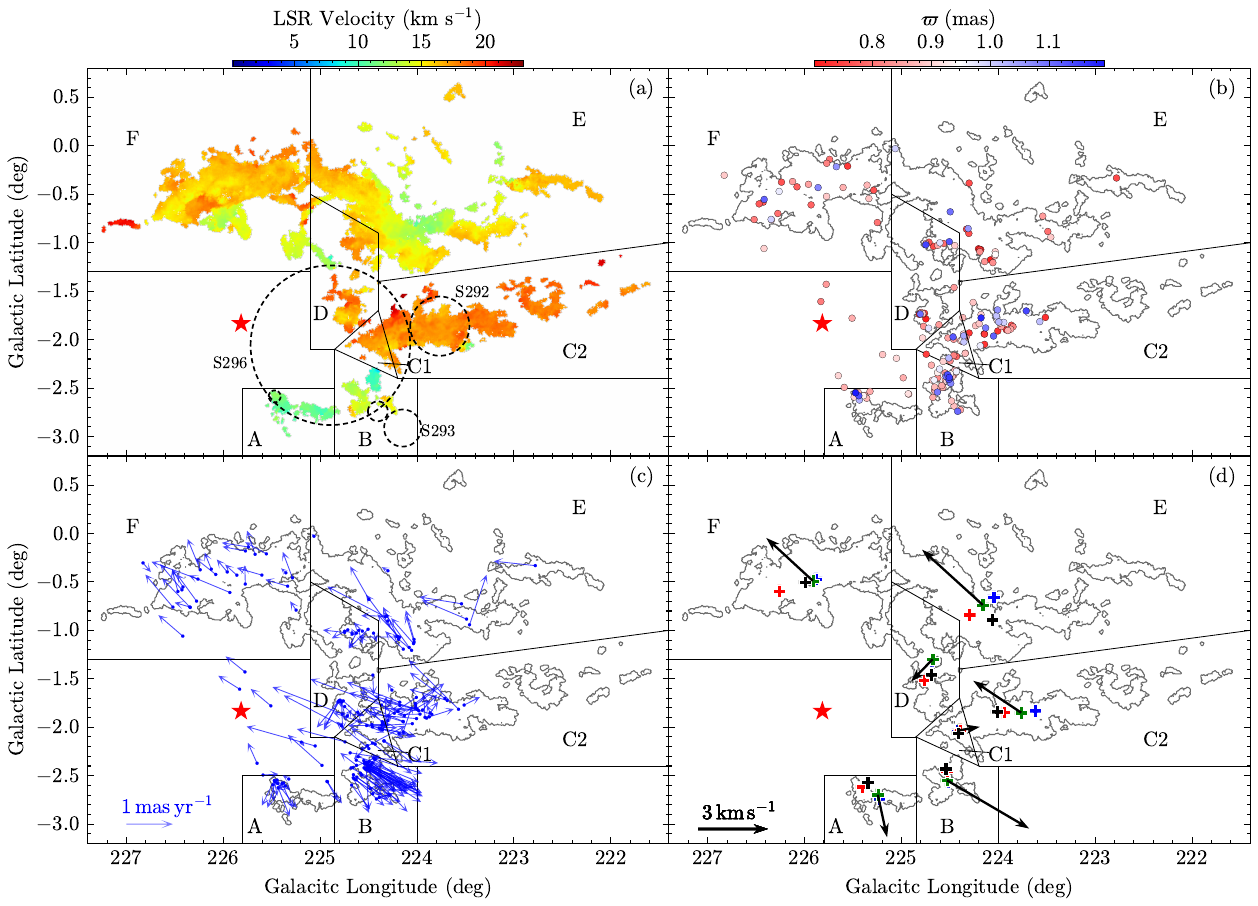}
	\caption{Division of subregions of CMa. Black solid lines delineate the boundaries of 7 subregions. Red stars denote the center of the CMa shell \citep{fernandes2019}. 
	(a): velocity distribution map of \lco. Black dashed circles mark \ion{H}{2} regions \citep[parameters from the WISE catalog;][]{Anderson_2014}. 
	(b): locations of YSOs. The YSOs are displayed as circles color-coded by their {parallaxes}. The background gray contours delineate the regions traced by \lco. 
	(c): motions of YSOs. Blue arrows represent residual tangential motions after subtracting the average motions of YSOs. 
	(d): the average positions of \uco, \lco, \ltco, and YSOs of subregions are shown as blue, green, red, and black crosses, respectively. Black arrows represent the average residual tangential velocities (converted to \kms) of YSOs within subregions.
	\label{fig:dividing}}
\end{figure*}

{The above studies provide compelling evidence that MCs in the CMa region consist of multiple components in the PPV space. 
As also indicated by these studies, different components may feature distinct dynamical states, and display varying levels of star formation activity. 
To date, there is no 3D view of morphology and motions for MCs in the CMa region. 
To initiate such a study, the first step is to resolve gas components with similar dynamical properties, i.e., division of subregions. 
Now, we can achieve this by integrating the MWISP CO data and YSO astrometry from \gaia~DR3.}

On the one hand, the CMa region can be divided based on the distributions of molecular gas. 
The two small clouds ({Clouds} 2 and 3) in the southern part appear as independent structures of the \uco~data in the PPV space. 
The giant MC ({Cloud} 1) in the north seems to be continuous in the PPV space.
Still, the integrated intensity map in Figure \ref{fig:m0} (a) shows that the molecular gas is not tightly connected in some areas of {Cloud} 1, suggesting the possible presence of separations. 
Figure \ref{fig:dividing} (a) presents the LSR velocity distribution of the relatively dense gas traced by \lco~(only showing the three studied MCs). 
The two southern clouds share the similar radial velocities that are smaller than the overall \vlsr~of the northern cloud.
Moreover, the gas near the \ion{H}{2} region S292 (black dashed circle) has significantly higher \vlsr~than the rest of the region.

On the other hand, we consider using the YSOs to facilitate the demarcation of subregions. 
In Figure \ref{fig:dividing} (b), the YSOs are marked as circles color-coded by their {parallaxes}, and the background gray contours depict the regions traced by \lco~of the three targeted MCs. 
{For Cloud 1, the overall parallax in the area of $b>-1.5\degree$ represents the smallest level in the CMa region, while in the area around $b\sim-2\degree$, the overall parallax remains relatively small. In contrast, YSOs in Clouds 2 and 3 exhibit relatively large overall parallaxes. This indicates that the YSOs in the northern region are generally more distant than those in the southern region.}
%

\begin{figure*}[t!]
	\centering
	\includegraphics[width=0.95\textwidth]{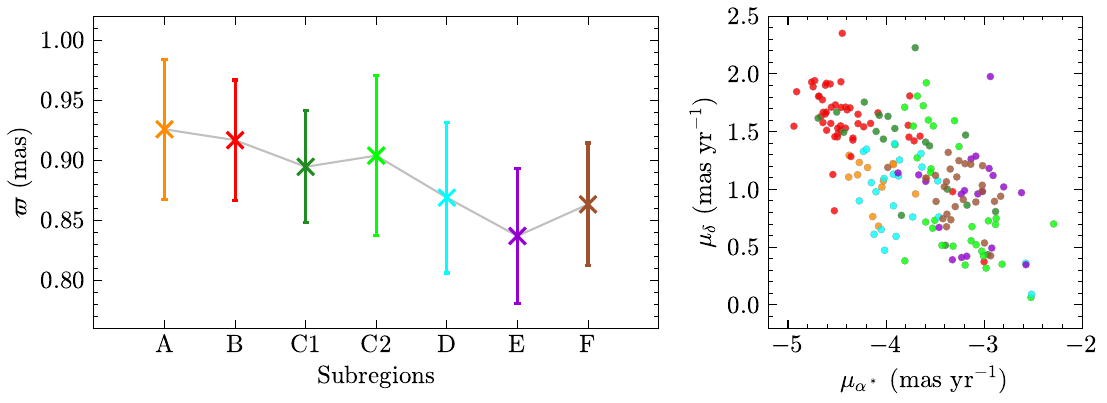}
	\caption{Parallaxes and proper motions of YSOs. Subregions A, B, C1, C2, D, {E, and F} are plotted in orange, red, dark green, light green, cyan, purple, and brown, respectively.
	{Left: average values (crosses) and standard deviations (vertical bars) of parallaxes.} 
	Right: distribution of YSOs in the proper motion space.
		\label{fig:subregion}}
\end{figure*}

Figure \ref{fig:dividing} (c) displays the residual tangential velocities of YSOs after subtracting the overall mean tangential velocity. 
The directions and lengths of the blue arrows represent the orientations and magnitudes of the residual tangential velocities. 
Clearly, the YSOs in different parts exhibit bulk motions {in diverse directions}. 
The YSOs located in the southern and northern regions move in almost opposite directions, while those located in the central region move in a more chaotic pattern, with some components moving towards the south and others moving along the east or west directions.

Taking into account the discrepancies in the PPV distribution, integrated intensity, and radial velocity of the molecular gas, along with the differences in {parallaxes} and motions of the YSOs, we decide to split the CMa region into 7 subregions, as indicated by the black solid lines in Figure \ref{fig:dividing}. 
Clouds 2 and 3 correspond to subregions A and B, respectively. 
Cloud 1 is divided into five subregions---C1, C2, D, E, and F. 
Note that subregions C1 and C2, although tightly connected in the CO data in PPV space, are separated based on the differences in the motions of YSOs. 
{For subregion D, the southern and northern parts exhibit comparable LSR velocities, parallaxes, proper motions, and residual tangential velocities, which are distinct from those observed in subregions C1 and E, respectively.}

The {average values} of parallaxes (left) and distributions of proper motions (right) are shown in Figure \ref{fig:subregion}, with the subregions plotted in different colors. 
The subregions not only demonstrate {the presence of parallax gradients}, but also are concentrated in distinct areas of the proper motion space.
This suggests that our division of the subregions is largely effective in distinguishing the different components of the CMa region. 
Histograms of YSOs parallaxes and proper motions in the subregions are presented as supplementary information in  Figure \ref{fig:histograms} in Appendix \ref{app:auxiliary}.

\subsection{6D Parameters of Subregions}\label{sec:6d_subregion}
Determining the six-dimensional (6D, 3D position and 3D motion) parameters of the subregions requires ($l$, $b$) positions, distances, proper motions (\pmra~and \pmdec), and radial velocities.
We {employ the average ($l$, $b$) coordinates of \lco, weighted by the main-beam brightness temperature ($ T\rm_{MB} $)}, to represent the centroid of the subregion, 
as \lco~traces denser regions of MCs than \uco~and has better signal-to-noise ratios than \ltco. 
Direct measurements of the distances and proper motions of the MCs solely from the CO observations are {not feasible}, so we supplement these data with the average parameters of the YSOs sample.
In particular, a large number of YSOs {allows} for more reliable and accurate determinations of distances and proper motions for the subregions. 
However, only a small percentage of YSOs (less than 10\%) in the CMa region have \gaia~radial velocity measurements, {and these are subject to large uncertainties}. 
Other spectroscopic surveys providing stellar radial velocities, such as LAMOST \citep{LAMOST}, 
RAVE \citep{RAVE}, and
APOGEE DR17 \citep{APOGEEdr17}, also lack observations towards the CMa region. 
Therefore, we adopt the $ T\rm_{MB} $-weighted average \vlsr~of \lco~as the radial velocities of the subregions. 
%

\setlength{\tabcolsep}{0.9mm}
\begin{deluxetable*}{cccccccccccchh}
	\tablecaption{Averaged Properties of Subregions
	\label{tab:ysos_co}}
	\tablehead{
		\colhead{Subregion} &\multicolumn{7}{c}{Class II or Earlier Class YSOs}&\multicolumn{6}{c}{Molecular Gas}\\
		\cmidrule(l{4pt}r{4pt}){2-8}
		\cmidrule(l{4pt}r{4pt}){9-13}
		\colhead{}
		&\colhead{$ N $}&\colhead{$ l $}&\colhead{$ b $} & \colhead{\plx} &\colhead{$ d\rm_{helio} $}&	\colhead{\pmra} &\colhead{\pmdec}&\colhead{$ l\rm_{13} $}&\colhead{$ b\rm_{13} $} & \colhead{$v\rm_{LSR,13}$} &	\colhead{{Mass}}&	\nocolhead{$ M\rm_{13} $}&	\nocolhead{$ M\rm_{18} $}\\
		\colhead{}&\colhead{}
		&\colhead{(\degree)}&\colhead{(\degree)}&
		\colhead{(mas)} &\colhead{(pc)} & \colhead{(\masyr)} 
		&\colhead{(\masyr)}&\colhead{(\degree)}&
		\colhead{(\degree)}&
		\colhead{(\kms)} & \colhead{({$\times10^3$ \msun})} & \nocolhead{($\times10^3$ \msun)} & \nocolhead{($\times10^3$ \msun)} 
	}
	\decimalcolnumbers
	\startdata
	A & 13 & 225.3 & $-2.6$ & $0.93\pm0.03$ & $1080\pm41$ & $-4.07\pm0.04$ & $1.08\pm0.05$ & 225.2 & $-2.7$ & 11.2 (0.8) & 2.9 & 1.5 & 0.16 \\
	B & {50} & 224.5 & $-2.4$ & $0.92\pm0.02$ & $1087\pm19$ & $-4.41\pm0.02$ & $1.62\pm0.02$ & 224.5 & $-2.5$ & 12.4 (2.1) & 3.4 & 2.4 & 0.27 \\
	C1 & 19 & 224.4 & $-2.1$ & $0.89\pm0.04$ & $1118\pm45$ & $-3.77\pm0.03$ & $1.32\pm0.03$ & 224.4 & $-2.0$ & 17.2 (0.9) & 4.3 & 3.6 & 1.37 \\
	C2 & 32 & 224.0 & $-1.8$ & $0.90\pm0.03$ & $1107\pm35$ & $-3.31\pm0.03$ & $0.94\pm0.03$ & 223.8 & $-1.8$ & 17.7 (1.1) & 28.4 & 17.7 & 2.30 \\
	D & 22 & 224.7 & $-1.5$ & $0.87\pm0.03$ & $1151\pm44$ & $-3.80\pm0.03$ & $0.93\pm0.03$ & 224.7 & $-1.3$ & 16.3 (1.5) & 8.6 & 4.2 & 0.46 \\
	E & 20 & 224.1 & $-0.9$ & $0.84\pm0.03$ & $1195\pm47$ & $-3.14\pm0.03$ & $0.94\pm0.03$ & 224.2 & $-0.7$ & 15.0 (1.3) & 37.8 & 18.7 & 4.83 \\
	F & 26 & 226.0 & $-0.5$ & $0.86\pm0.03$ & $1159\pm38$ & $-3.26\pm0.03$ & $0.97\pm0.03$ & 225.9 & $-0.5$ & 16.0 (1.3) & 26.0 & 10.7 & 0.60 \\
	\enddata
	\tablecomments{
	{Column 2 lists the number of YSOs. Columns 3--8 provide the averaged properties estimated from YSOs.} 
	The uncertainties in parallaxes, distances, and proper motions in Columns 5--8 are derived from the Monte Carlo simulations considering the \gaia's observational errors. 
	Columns 9--11 represent the $ T\rm_{MB} $-weighted average $l$, $b$, and \vlsr~of the \lco~line. 
	The velocity dispersions in parentheses in Column 11 are estimated as the $ T\rm_{MB} $-weighted standard deviations of the \lco-traced \vlsr, {serving as the uncertainties}.
	{Column 12 lists the mass of molecular gas traced by \uco, recalculated using the distance in Column 6.}
	}
\end{deluxetable*}

Table \ref{tab:ysos_co} presents the observational parameters for different subregions in CMa. 
{The parameters of YSOs are calculated after excluding outliers, whose parallaxes and proper motions deviate by more than 3$\sigma$ from their respective average values. 
In Column 6, the distances are determined by inverting the mean parallaxes of YSOs, following the method used in many previous studies \citep[e.g.,][]{orion2018,orion2021,gh21}.} 
The average distances of 7 subregions are estimated to be {$1128\pm15$} pc, which is basically in accordance with the distance measurements of the CMa clouds in the literature, such as {\citet[1150 pc]{Kim_2004}},
\citet[$ 1150\pm64 $ pc]{Lombardi2011}, 
\citet[$1209\pm60 $ pc]{zucker2019}, 
\citet[$ 1185\pm25 $ pc]{PR19}, 
\citet[$\sim$1150 pc]{Dharmawardena2023}, and 
\citet[$1057^{+29}_{-27}$ pc]{z23}.

Meanwhile, the distance gradient of $ \sim$100 pc is revealed, with subregion A being the closest ($ \sim$1080 pc), subregion E the most distant ($ \sim$1195 pc), and other subregions lying in between. 
In Column 11, differences in the radial velocities of the subregions are pronounced, with subregions A and B having radial velocities $\sim$4--5 \kms~lower than subregions C1--F. 
Further, we convert the proper motions of the subregions into the residual tangential velocities in the sky plane in units of \kms, drawn as black arrows in Figure \ref{fig:dividing} (d), to provide a clear picture of the global motions.
Obviously, the southern and northern subregions move towards almost opposite directions, suggesting a potential expanding pattern for the entire region.
Subsequently, we utilize the parameters in Table \ref{tab:ysos_co} to derive the 3D spatial coordinates and 3D velocities of the subregions in the Milky Way. 
The constructed Galactic Cartesian coordinate system ($ o-xyz $) and the fundamental Galactic and solar parameters refer to \citet[model fit A5]{Reid_2019}. 
The results are presented as supporting information in Table \ref{tab:3d} in Appendix \ref{app:auxiliary}.

\subsection{3D Morphology and Motions of the CMa Region} \label{sec:3D}
\begin{figure*}[t!]
	\centering
	\gridline{\fig{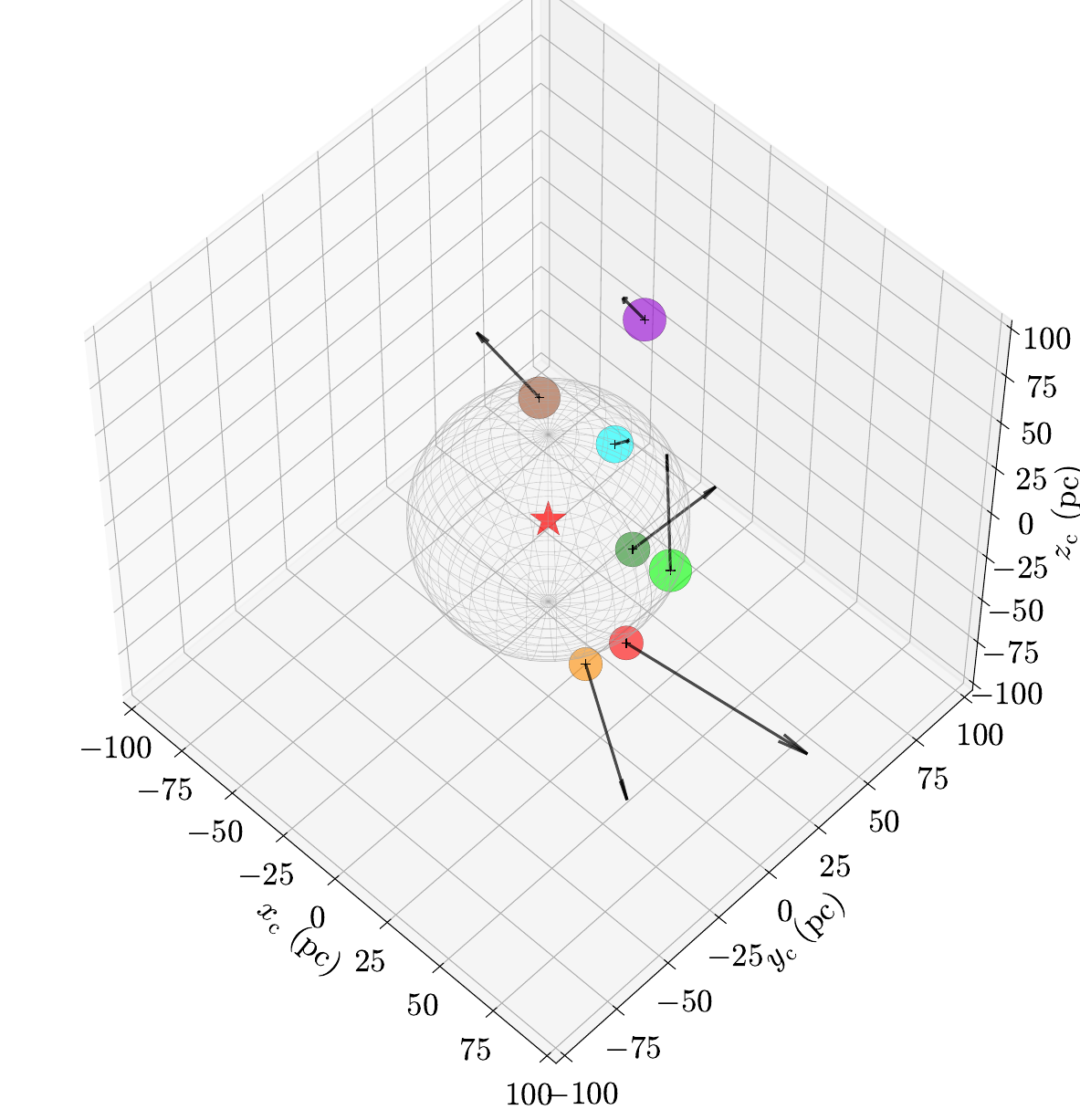}{0.5\textwidth}{}}
	\vspace{-0.8cm}
	\gridline{\fig{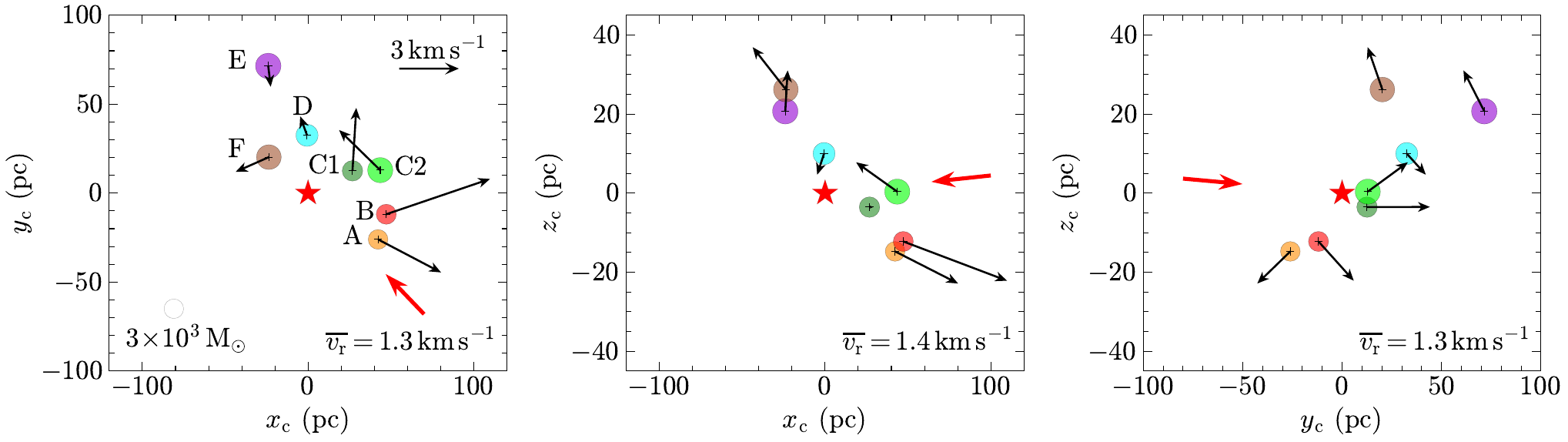}{1\textwidth}{}}
	\vspace{-0.8cm}
	\caption{3D distributions and motions of the subregions. 
	The red star at the origin of the coordinate system is the center of the CMa shell, and the positive directions of the $ x_{\rm c} $, $ y_{\rm c} $, and $ z_{\rm c}$ axes are consistent with the Galactic Cartesian coordinate system of \citet{Reid_2019}. 
	The positions and motions of the subregions are marked with colored circles and black arrows, respectively. The circle colors are aligned with Figure \ref{fig:subregion}, and the circle sizes are proportional to the logarithm of the molecular masses of the subregions {(listed in Table \ref{tab:ysos_co})}. 
	Upper panel: the 3D view, with gray grid lines depicting a spherical shell with a radius of $\sim${47} pc. 
	Lower panels: projected views on the $ x_{\rm c}$--$y_{\rm c} $ (left), $ x_{\rm c}$--$z_{\rm c} $ (middle), and $ y_{\rm c}$--$z_{\rm c} $ (right) planes. 
	Red arrows indicate the line-of-sight direction. 
	The averaged radial components of velocities of the subregions are labeled in the bottom-right corners, with positive values pointing outwards and negative values pointing inwards.
	\label{fig:3d_project}}
\end{figure*}

In conjunction with the distances and proper motions of the YSOs, the CMa region exhibits a large-scale molecular structure with internal relative motions.
With the aim of establishing its 3D morphology and motions, we select an appropriate reference point to calculate the relative positions and velocities of the subregions. 
Here, the CMa shell center proposed by \citet{fernandes2019} is employed as the reference point ($(l,~b)=(225.82\degree,\,-1.83\degree)$, see the red stars in Figures \ref{fig:m0}, \ref{fig:dividing}, and subsequent figures). 
The distance of the shell center is the aforementioned average distance ({$1128\pm15$} pc), and the \pmra, \pmdec, and \vlsr~of the shell center are obtained from the average parameters of the 7 subregions, yielding values of {$-3.68\pm0.01$} \masyr, {$1.11\pm0.01$} \masyr, and {$15.1\pm0.5$} \kms, respectively. 
After that, we can calculate the 3D positions and motions of the subregions relative to the shell center.
These values, listed in Table \ref{tab:3d_rel} and visualized in Figure \ref{fig:3d_project}, are represented in the Cartesian coordinate system ($ o_{\rm c}-x_{\rm c}y_{\rm c}z_{\rm c} $), where the origin $ o_{\rm c}$ is the shell center, and the $ x\rm_{c}$, $ y\rm_{c}$, and $ z\rm_{c}$ axes align with the axes of the Galactic Cartesian coordinate system.
%

\begin{deluxetable*}{chrrrrrrrrrrr}
	\tablecaption{{Summary of Positions and Motions of Subregions}
	\label{tab:3d_rel}}
	\tablehead{
		\colhead{Subregion} 
		&\nocolhead{$ d\rm_{helio} $}	&\colhead{$ x\rm_{c}$}&\colhead{$ y\rm_{c} $} & \colhead{$ z\rm_{c}$} &	\colhead{$v_{x\rm_{c}} $}&	\colhead{$v_{y\rm_{c}} $}&	\colhead{$v_{z\rm_{c}} $}&	\colhead{$\vert\boldsymbol{r}\vert$}&\colhead{$\hat{\boldsymbol{r}}\cdot\boldsymbol{v}$}&\multicolumn{3}{c}{$\hat{\boldsymbol{r}}\times \boldsymbol{v}$}\\ 
		 \cmidrule(l{4pt}r{4pt}){11-13}
		\colhead{}&\nocolhead{(pc)} &\colhead{(pc)}&
		\colhead{(pc)}&
		\colhead{(pc)}&
		\colhead{(\kms)}&
		\colhead{(\kms)}&
		\colhead{(\kms)}&
		\colhead{(pc)}&
		\colhead{(\kms)}&
		\multicolumn{3}{c}{(\kms)}
	}
	\decimalcolnumbers
	\startdata
		A & $1080\pm41$ & $42.3$ & $-26.0$ & $-14.7$ & $3.2\quad$ & $-1.7\quad$ & $-1.6\quad$ & $51.8$ & $3.9\quad$ & $0.3$ & $0.4$ & $0.2$ \\
		B & $1087\pm19$ & $47.2$ & $-11.9$ & $-12.2$ & $5.2\quad$ & $1.8\quad$ & $-2.0\quad$ & $50.2$ & $5.0\quad$ & $0.9$ & $0.6$ & $2.9$ \\
		C1 & $1118\pm44$ & $26.8$ & $12.5$ & $-3.5$ & $0.2\quad$ & $3.2\quad$ & $0.01\quad$ & $29.8$ & $1.5\quad$ & $0.4$ & $-0.03$ & $2.8$ \\
		C2 & $1107\pm35$ & $43.6$ & $12.9$ & $0.4$ & $-2.1\quad$ & $2.0\quad$ & $1.5\quad$ & $45.4$ & $-1.4\quad$ & $0.4$ & $-1.4$ & $2.5$ \\
		D & $1151\pm44$ & $-0.6$ & $32.5$ & $10.0$ & $-0.3\quad$ & $1.0\quad$ & $-1.1\quad$ & $34.0$ & $0.6\quad$ & $-1.3$ & $-0.1$ & $0.3$ \\
		E & $1195\pm47$ & $-24.0$ & $71.5$ & $20.7$ & $0.1\quad$ & $-1.0\quad$ & $2.1\quad$ & $78.2$ & $-0.5\quad$ & $2.2$ & $0.7$ & $0.2$ \\
		F & $1159\pm39$ & $-23.7$ & $20.2$ & $26.2$ & $-1.7\quad$ & $-0.7\quad$ & $2.1\quad$ & $40.7$ & $2.0\quad$ & $1.5$ & $0.2$ & $1.3$ \\
	\enddata
	\tablecomments{Columns 2--7: 3D positions and motions of the subregions relative to the center of the CMa shell. 
	Column 8: distances of the subregions to the shell center.
	Column 9: dot products for the position vectors (defined by Columns 2--4) and velocity vectors (defined by Columns 5--7), as measures of expansion.
	Columns 10--12: cross products for the position vectors and velocity vectors, as measures of rotation.
	}
\end{deluxetable*}

\subsubsection{3D Morphology} \label{sec:3Dstructure}
\citet{fernandes2019} proposed a shell with a diameter of $\sim$60 pc based on multi-band 2D projection images of the CMa region; we now seek to investigate its 3D morphology. 
From Figure \ref{fig:3d_project}, we observe that the subregions seem to be distributed around a 3D spherical structure. 
In the three projected views in the lower panels of Figure \ref{fig:3d_project}, with the line-of-sight direction indicated by red arrows, it is apparent that subregions A, B, C1, and C2 are located on the front wall of the shell (closer to the Sun), while subregions D, E, and F are situated on the back wall of the shell (further from the Sun). 
To measure the size of the shell, we denote the 3D coordinates of the subregions using the position vectors $\boldsymbol{r}=(x_{\rm c},~y_{\rm c},~z_{\rm c})$. 
The distance from each subregion to the shell center equals $\vert\boldsymbol{r}\vert$, as listed in the Column 8 of Table \ref{tab:3d_rel}. 
Averaging $\vert\boldsymbol{r}\vert$ of the 7 subregions, we determine the radius of the shell to be {$47\pm11$} pc, which represents a 3D shell larger than the model of \citet{fernandes2019}. 
The shell is plotted as an imaginary image in the upper panel of Figure \ref{fig:3d_project}. 

We find that the 3D distributions of the 7 subregions can be categorized into three situations: 
(1) Subregions A, B, C2, and F appear to be located at the rim of the shell, as their distances to the shell center are similar to the radius of the shell; 
(2) Subregions C1 and D are situated on the inner side of the shell, with distances to the shell center smaller than the radius of the shell; 
(3) Subregion E is significantly further from the shell, with a distance to the shell center almost twice the radius of the shell.
{Note that the larger shell radius compared to the estimate of \citet{fernandes2019} is not due to the higher $\vert\boldsymbol{r}\vert$ of subregion E. 
Even if subregion E is excluded, the average $\vert\boldsymbol{r}\vert$ is still $\sim$42 pc, representing a larger shell.}

\subsubsection{3D Motions} \label{sec:3Dmotion} 
The results in Section \ref{sec:6d_subregion} suggest that the CMa region may be undergoing expansion, a trend also indicated by the black arrows in Figure \ref{fig:3d_project}. 
In this section, we analyze the 3D motions of the subregions.
Specifically, for each of the three projected views in the lower panels of Figure \ref{fig:3d_project}, we compute the average radial component ($\overline{v_{\rm r}}$) of the velocities of subregions relative to the shell center, as labeled in the lower-right corner of the plot.
These positive $\overline{v_{\rm r}}$ values collectively demonstrate an overall outward radial motion, signifying the presence of expansion on all three projection planes.

To eliminate the influence of projection effects, we measure the expansion with radial motions in 3D space. 
Following the method described by \citet{rivera2015}, we employ $\hat{\boldsymbol{r}}\cdot \boldsymbol{v}$ to characterize the expansion (positive value) or contraction (negative value) motions, where $\hat{\boldsymbol{r}}=\boldsymbol{r}/\vert\boldsymbol{r}\vert$ are the unit vector of the position vector, and $\boldsymbol{v}=(v_{x\rm_{c}},~v_{y\rm_{c}},~v_{z\rm_{c}})$ denotes the velocity vector of each subregion.
In fact, $\hat{\boldsymbol{r}}\cdot \boldsymbol{v}$ represents the radial component of the 3D velocities, as listed in Column 9 in Table \ref{tab:3d_rel}.
The average $\hat{\boldsymbol{r}}\cdot \boldsymbol{v}$ for the 7 subregions is {$1.6\pm0.7$} \kms, a value comparable to the velocity dispersions of MCs ($\sim$2.2, 1.9, and 1.8 \kms~for \multilines, respectively).
This suggests that the global expansion of the molecular shell is unlikely to be dominated by cloud turbulence, but is likely to be attributed to the external force. 

Incidentally, we calculate $\hat{\boldsymbol{r}}\times \boldsymbol{v}$ (see Columns 10--12 in Table \ref{tab:3d_rel}) as a proxy for the rotational signal;  
however, it is important to note that $\hat{\boldsymbol{r}}\times \boldsymbol{v}$ is not strictly the rotation velocity.
The average $\hat{\boldsymbol{r}}\times \boldsymbol{v}$ for the 7 subregions are 0.6, 0.03, and 1.5 \kms, which appear to be insignificant compared to the velocity dispersions. This indicates the absence of a discernible organized rotation pattern of the CMa molecular shell.

\section{Discussion} \label{sec:discussion}
\subsection{Gas Column Density Profile of Expanding Shell}\label{sec:gas_columndensity}
Previous studies have shown that the molecular material can display clearly asymmetric gas column density profiles due to the compression by the feedback force, i.e., the compressed side exhibits a sharp edge \citep[e.g.,][]{peretto2012,schneider2013,zavagno2020}.
The prevailing view suggests that the shell-like structure in the CMa region is shaped by an old SNR (e.g., \citealt{gh2008} and references therein). 
The question arises whether the molecular gas in the CMa region is compressed during the SNR expansion. 
Our objective is to utilize the column density profiles of the molecular gas to investigate the potential clues of the expanding SNR.

\begin{figure*}[t!]
	\centering
	\gridline{\fig{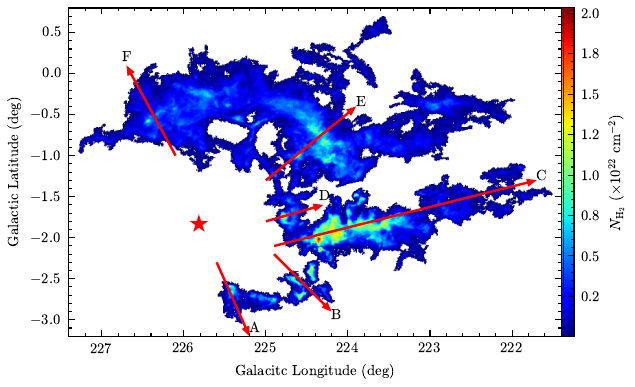}{0.6\textwidth}{}
	\fig{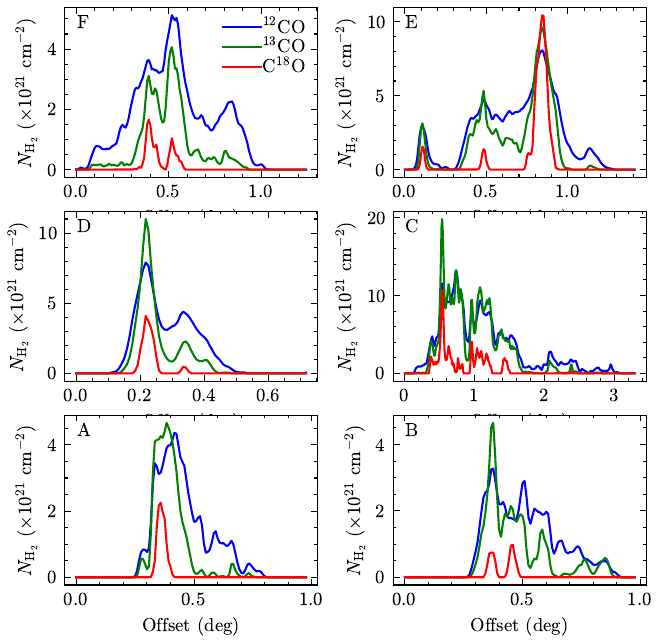}{0.38\textwidth}{}
	}
	\vspace{-0.7cm}
	\caption{Gas column density features of the CMa region. Left panel: distribution of $N_{\rm H_2}$ traced by \uco, with the center of the CMa shell marked by a red star. Red arrows represent the 6 directions for P-$N_{\rm H_2}$ slicing. Right panels: column density profiles along the 6 directions, arranged in the order of the arrows in the left panel, with the blue, green, and red lines depicting $N_{\rm H_2}$ traced by \multilines, respectively.
			\label{fig:pNH2}}
\end{figure*}

The left panel of Figure \ref{fig:pNH2} displays the column density of three MCs traced by \uco~(see \citetalias{Dong2023} for the details of calculation), with the center of the CMa shell marked by the red star. 
We extract six position-$\rm H_2$ column density (P-$N_{\rm H_2}$) profiles that cross the subregions roughly in the radial directions relative to the shell center, as indicated by the red arrows in Figure \ref{fig:pNH2}. 
The arrows are designated by subregions, with arrow C representing both subregions C1 and C2. 
The P-$N_{\rm H_2}$ profiles are presented in the right panels of Figure \ref{fig:pNH2}, arranged according to the positions of arrows in the left panel.
Blue, green, and red profiles depict $N_{\rm H_2}$ traced by \multilines, respectively.

Evidently asymmetric distributions of gas column densities are observed along directions A, B, C, and D, with steep edges on the inner side of the shell (positions with small offsets). 
The gas is notably thinner in direction F, where the denser molecular gas tracers \lco~and \ltco~exhibit steep column density profiles on the inner side of the shell. 
However, the P-$N_{\rm H_2}$ profiles in direction E show no obvious asymmetry, possibly due to the fact subregion E is situated far away from the shell and exposed to weaker compressive effects. 
This, in turn, supports the validity of dividing subregions based on the 3D positions and motions, and highlights the importance of 3D studies. 
In conclusion, the column density profiles {confirm} that the {shell-like molecular structure} of the CMa region is subjected to compression from the interior, with the expanding SNR being a plausible source of the external force {(e.g., \citealt{gh2008} and references therein)}.

\subsection{Possible Origin of Expanding Shell}\label{sec:evolve}
{Our investigation into the 3D morphology and kinematics of the CMa region, 
integrating molecular gas data and YSO astrometry, corroborates the expanding shell in the CMa region previously identified by \ion{H}{1} and H$\alpha$ observations \citep{Herbst1977,reynolds1978}.} 
In this section, we elucidate {the} dynamical evolution and potential driving mechanism {behind the expansion of the molecular material within the shell}.
To pinpoint the time when the 7 subregions are most compactly distributed, 
we trace their past and future trajectories (within a time range of $\pm$20 Myr) following the method of \citet{orion2021}.

\begin{figure}[t!]
	\centering
	\includegraphics[width=0.45\textwidth]{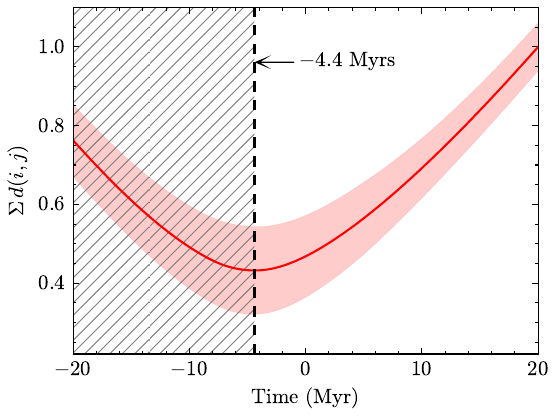}
	\caption{The sum of the distances between the 7 subregions $\Sigma\,d{(i,j)}$ as a function of evolutionary time, normalized by its maximum value. {The red shaded area represents the standard deviations of the Monte Carlo sampling}. The time range is $\pm20$ Myr with a step size of 0.1 Myr. The vertical dashed line marks the moment when $\Sigma\,d{(i,j)}$ reaches a minimum.
		\label{fig:traceback}}
\end{figure}

Concretely speaking, we assume constant cloud velocities to calculate the sum of distances $\Sigma\,d{(i,j)}$ between the subregions at 0.1 Myr time bins and determine the minimum value. 
Figure \ref{fig:traceback} illustrates $\Sigma\,d{(i,j)}$ as a function of the evolutionary time, indicating that the expansion of the molecular shell seems to have started $\sim$4.4 Myr ago. 
{This dynamical age should be considered a rough estimate, since we are currently unable to provide a precise value or its associated uncertainty. 
As indicated by \citet{orion2021}, the limitation arises from the complexity of uncertainties involved in the tracebacks, such as measurement errors in the \gaia~data and CO data, and statistical biases.}

The evolutionary positions of the subregions within $\pm$4.4 Myr in the $ x_{\rm c}-y_{\rm c} $, $ x_{\rm c}-z_{\rm c} $, and $ y_{\rm c}-z_{\rm c} $ projected planes are shown in Figures \ref{fig:trajectory1}--\ref{fig:trajectory3} in Appendix \ref{app:auxiliary}, respectively. 
One can see that the subregions are indeed more concentrated and closer to the shell center before 4.4 Myr. 
Moving forward in time, the subregions become more separated from each other and gradually disperse in the 3D space, further confirming the expansion of the CMa region.

It is important to note that several factors can influence the inference of trajectories. 
The current estimates are based on the assumption of simple constant velocities, whereas in reality, various mechanisms can cause changes in velocities. 
For instance, multiple feedback events reinject momentum and energy into the shell to accelerate its expansion;  
the interactions of the expanding shock with the clouds result in a loss of kinetic energy and a gain of thermal energy, leading to the deceleration of the expansion;  
the gravitational interactions between clouds and the gravitational potential of the Milky Way may also alter the motions of the subregions (e.g., \citealt{chevalier1974}, \citealt{krause2014}, \citealt{borkowski2017}). 
Therefore, it is challenging to accurately determine the initial velocity of the molecular shell and the velocity changes during the expansion process. 
The expansion timescale we obtained (4.4 Myr) is smaller than the earliest time of SNe events reported by \citet[6 Myr, as also described in Section \ref{sec:intro}]{fernandes2019}. 
The discrepancy may be ascribed to multiple SNe events continuously releasing energy and momentum into the shell, increasing the expansion velocity, and thus resulting in a smaller dynamical age.

Next, we use momentum estimates to infer the number of SNe events required to drive the cloud motions \citep[e.g.,][]{orion2021,posch2023}. 
{According to Table \ref{tab:ysos_co}, the total mass of the clouds is $1.1\times10^5$ \msun.}  
With an expansion velocity of $\sim$1.6 \kms, the momentum of expansion is estimated as $\sim$$1.8\times10^5$ \msun\,\kms. 
According to simulations, a single SNe event typically outputs momentum of $\sim$2--$4\times10^5$ \msun\,\kms~\citep[e.g.,][]{iffrig2015,kim2015_SNe,walch2015}. 
Based on the traceback results, the radius of the shell at 4.4 Myr ago is {$R_0\sim41$} pc, from which we can derive the surface area of the shell ($A_{\rm shell}=4\pi R_0^2$). 
The current projected area of the cloud on the sky plane is about $a_{\rm now}\sim2.6\times10^3\rm~pc^2$. 
Referring to \citet{orion2021}, we test half to twice as large as the current projected area subjected to the driving force of the SNe.
To calculate the lower limit of the number of SNe, we assume that all the momentum from the SNe in this portion of the surface area, 0.5--2 $a_{\rm now}$, has been transferred to the MCs.
Therefore, it requires at least 2 SNe events to power the expanding molecular shell. 
Indeed, \citet{fernandes2019} argued that at least 3 SNe events may have occurred in this region based on the trajectories of runaway stars. 
However, we note that this rough estimate overlooks potential feedback mechanisms before SNe such as the stellar winds and radiation pressure from massive stars, which could also contribute to the observed scenario.

\section{Conclusions} \label{sec:conclusions}Integrating the high-precision astrometric data of YSOs from \gaia~DR3 with the high-quality \multilines~(1–0) data from the MWISP project, we investigate for the first time 
the 3D morphology and motions of the CMa region. 
To initiate this study, the CMa region is partitioned into 7 subregions, 
taking into account the discrepancies in the PPV distribution, integrated intensity, and radial velocity of the molecular gas, along with the differences in distances and motions of the YSOs. 
The central ($l$, $b$) coordinates and radial velocities of the subregions are determined by the $ T\rm_{MB} $-weighted average parameters of the \lco~gas, and the distances and proper motions of the subregions are represented by the average parameters of the YSOs. 
These parameters enable us to infer the positions and motions of the subregions in the 3D physical space.

We find that the subregions are distributed around a shell with a radius of {$47\pm11$} pc. 
The radial components {(relative to the shell center)} of the 3D motions indicate that this molecular shell is slowly expanding at a velocity of {$1.6\pm0.7$} \kms.
{These results confirm previous findings based on 2D observations regarding the presence of expanding shell and its possible formation mechanisms \citep[e.g.,][]{Herbst1977,reynolds1978,machnik1980,fernandes2019}.}
Through the traceback analysis assuming constant velocities, we observe that the expansion appears to have started $\sim$4.4 Myr ago. 
The evolving trajectories of the subregions lend further support to the expanding motions. 
Additionally, the momentum analysis suggests that the expanding molecular shell may be shaped by at least 2 SNe events. 
The asymmetric radial gas column density profiles support the idea that the molecular shell is being compressed from the interior.
Finally, based on the available data and findings, we infer the large-scale evolutionary sequence of the CMa region, showing that the region has experienced multiple episodes of star formation, likely triggered by several SNe events, thus reinforcing the model proposed by \citet{fernandes2019}.

To conclude, this study provides new insights into the morphology and motions of the MCs in CMa region, giving important clues for understanding its dynamical evolution and star formation history. 
In the future, the ongoing MWISP project will cover a larger area of the Milky Way, and the upcoming \gaia~Data Releases will offer more elaborate astrometric measurements of YSOs. 
We look forward to continuing to combine MWISP and \gaia~data to explore the 3D morphology and kinematics of in MCs in different star-forming environments, gaining new insights into the interplay between stellar feedback and interstellar molecular materials.

\begin{acknowledgments}
We thank the referee, for insightful feedback that helped to improve the paper. 
This work was funded by the NSFC grant 11933011, the National SKA Program of China (grant No.\,2022SKA0120103), and the Key Laboratory for Radio Astronomy.  
C.J.H. acknowledges support from the National Postdoctoral Program for Innovative Talents of the Oﬀice of China Postdoc Council (grant No.\,BX20240414) and the NSFC grant 12403041. 
Y.J.L. thanks support from the NSFC grant 12203104 and the Natural Science Foundation of Jiangsu Province (grant No.\,BK20210999).
Y.S. acknowledges support by the Youth Innovation Promotion Association, CAS (Y2022085), and the ``Light of West China'' Program (No.\,xbzg-zdsys-202212).
This research used the data from the Milky Way Imaging Scroll Painting (MWISP) project, which is a multiline survey in \uco/\lco/\ltco~along the northern Galactic plane with the PMO 13.7 m telescope. MWISP was sponsored by the National Key R\&D Program of China with grant 2017YFA0402701 and the CAS Key Research Program of Frontier Sciences with grant QYZDJ-SSW-SLH047.
This work has made use of data from the European Space Agency (ESA) mission
{\it Gaia} (\url{https://www.cosmos.esa.int/gaia}), processed by the {\it Gaia}
Data Processing and Analysis Consortium (DPAC,
\url{https://www.cosmos.esa.int/web/gaia/dpac/consortium}). Funding for the DPAC
has been provided by national institutions, in particular the institutions
participating in the {\it Gaia} Multilateral Agreement.
This work has also made use of the VizieR, SIMBAD, and Aladin databases operated at CDS, Strasbourg, France.
\end{acknowledgments}

\software{Astropy \citep{Astropy2013,Astropy2018}, matplotlib \citep{Hunter_2007}, NumPy \citep{Numpy}}, SAOImage DS9 \citep{ds9}, TOPCAT \citep{TOPCAT}.

\appendix
\section{Auxiliary Figures and Tables} \label{app:auxiliary}
\begin{figure*}[ht!]
	\centering
	\figurenum{A1}
	\includegraphics[width=0.9\textwidth]{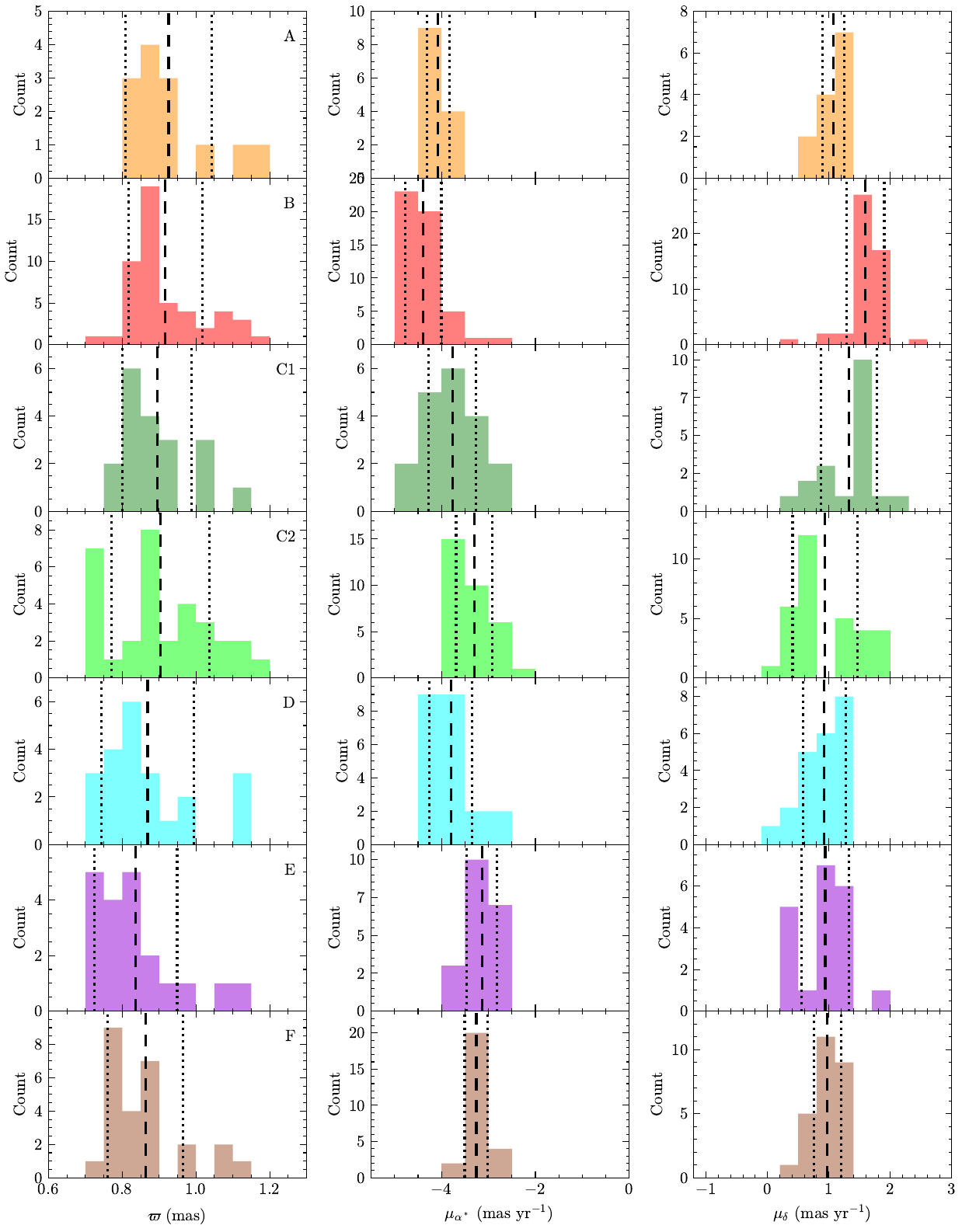}
	\caption{
	Histograms of the \plx~(left), \pmra~(middle), and \pmdec~(right) for the YSO samples in the subregions. 
	From top to bottom: the histograms of subregions A--F, with colors same as in Figure \ref{fig:subregion}. 
	Vertical dashed lines indicate the average values of the parameters, and dotted lines mark the positions of $\pm 1\sigma$ around the average values.
	\label{fig:histograms}}
\end{figure*}

The descriptions of the auxiliary figures and tables follow the order of their appearance in the main text. 
Figure \ref{fig:histograms} displays the histograms of the parallax (\plx) and proper motions (\pmra~and \pmdec) of the YSOs sample in the subregions. 
The parameters for subregions A--F are presented in order from top to bottom, in the same colors as Figure \ref{fig:subregion}. 
Vertical dashed lines mark the average values of the parameters.
Vertical dotted lines indicate the positions of $\pm 1\sigma$ around the average values.

Table \ref{tab:3d} lists the 3D coordinates and motions of the CMa shell center and the subregions in the Milky Way.
The derivation of these parameters is described in the main text.

\begin{figure*}[t!]
	\centering
	\figurenum{A2}
	\includegraphics[width=0.95\textwidth]{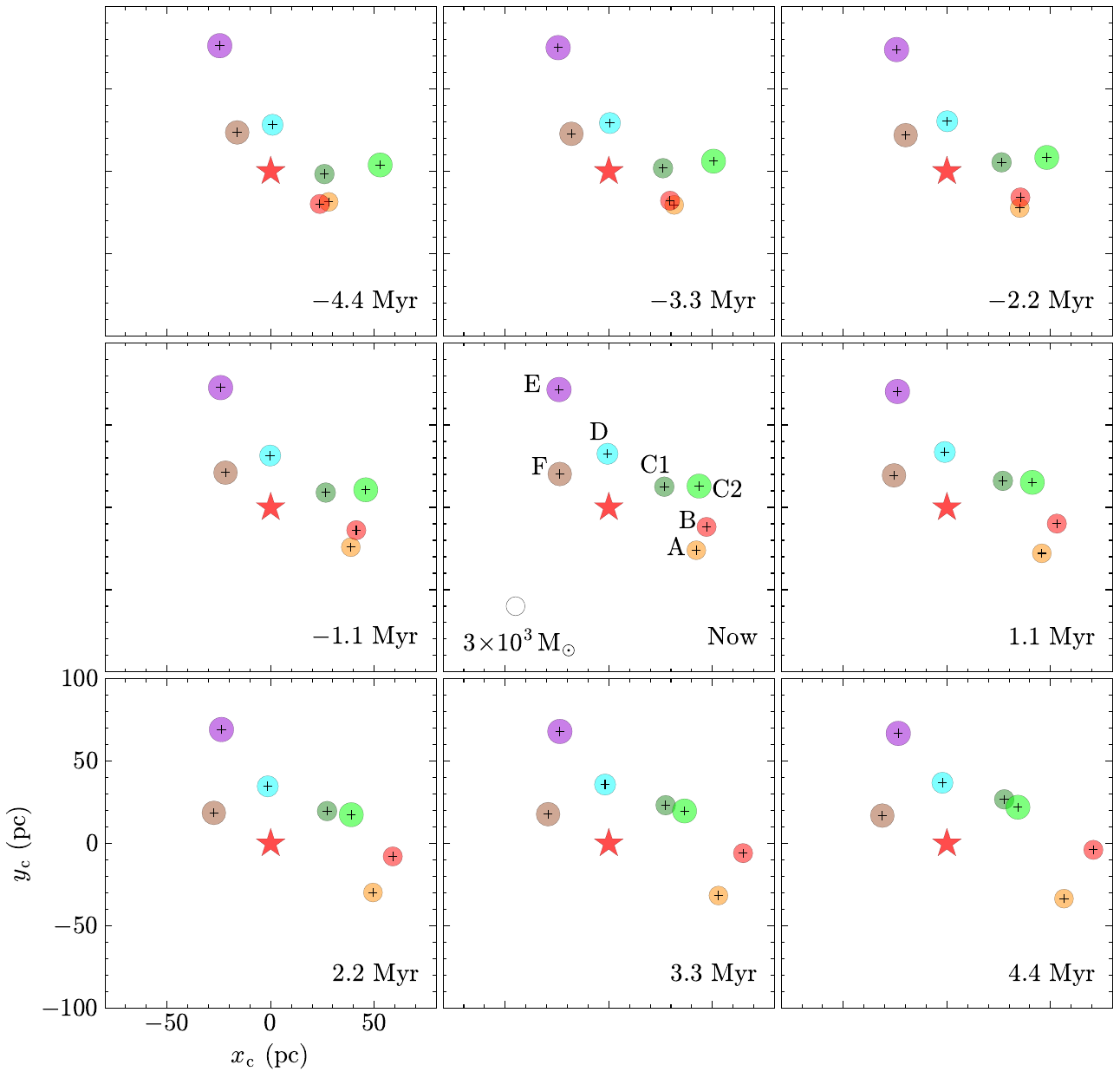}
	\caption{The distribution of the subregions within $\pm4.4$ Myr, projected onto the $x_{\rm c}-y_{\rm c}$ plane. The markers of the subregions are the same as in Figure \ref{fig:3d_project}.
		\label{fig:trajectory1}}
	\end{figure*}

\begin{figure*}[ht!]
	\centering
	\figurenum{A3}
	\includegraphics[width=0.95\textwidth]{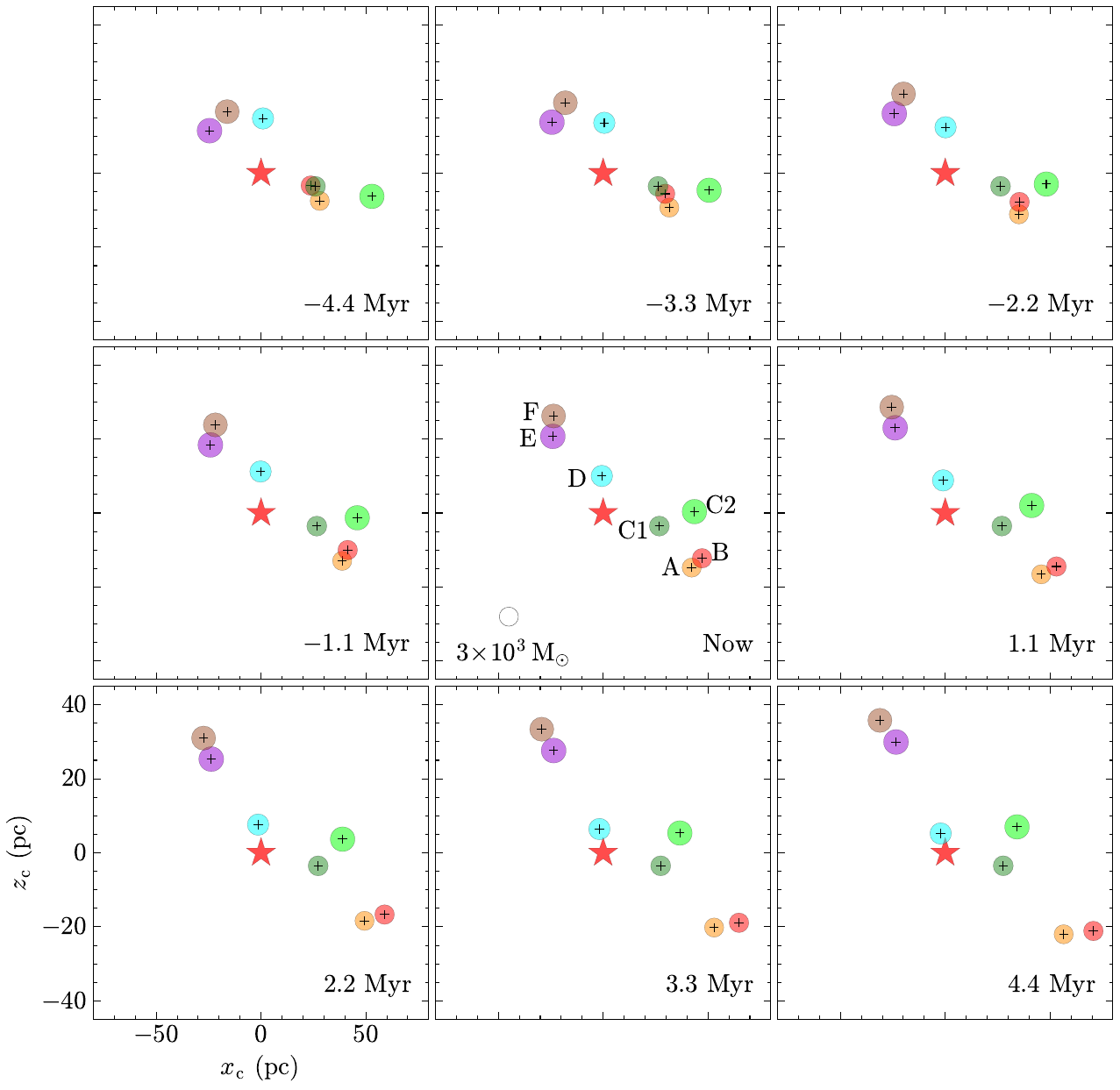}
	\caption{Same as Figure \ref{fig:trajectory1}, but projected onto the $x_{\rm c}-z_{\rm c}$ plane.
	\label{fig:trajectory2}}
	\end{figure*}

\begin{figure*}[ht!]
	\centering
	\figurenum{A4}
	\includegraphics[width=0.95\textwidth]{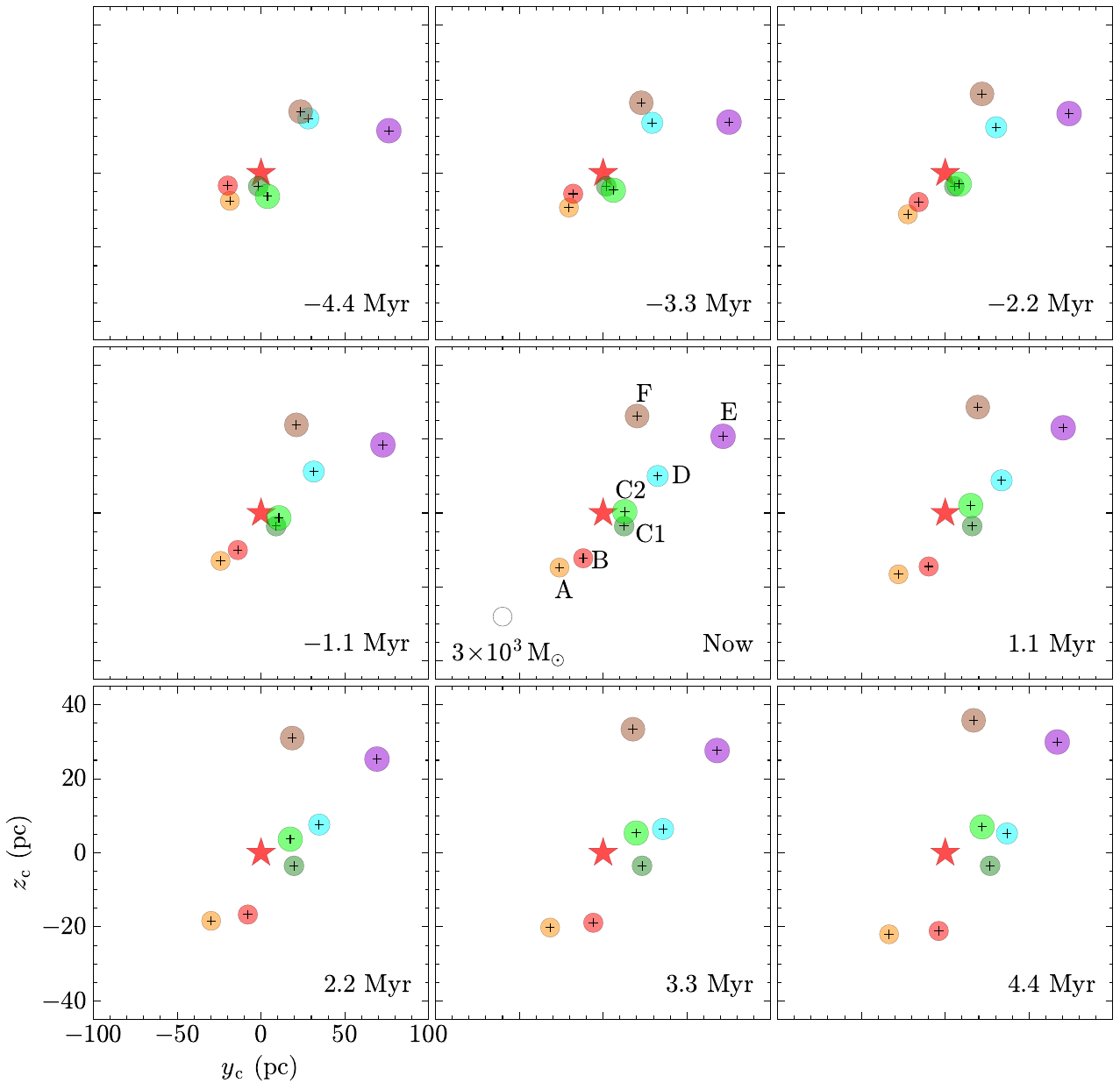}
	\caption{Same as Figure \ref{fig:trajectory1}, but projected onto the $y_{\rm c}-z_{\rm c}$ plane.
		\label{fig:trajectory3}}
	\end{figure*}

Figures \ref{fig:trajectory1}--\ref{fig:trajectory3} present the position evolutions of the subregions within $\pm4.4$ Myr, projected on the $x_{\rm c}-y_{\rm c}$, $x_{\rm c}-z_{\rm c}$, and $y_{\rm c}-z_{\rm c}$ planes, respectively.
The markers of subregions refer to Figure \ref{fig:3d_project}.
%

\begin{deluxetable*}{chrrrrrr}
	\tablenum{A1}
	\tablecaption{3D Positions and Motions in the Milky Way
	\label{tab:3d}}
	\tablehead{
		\colhead{Subregion} 
		&\nocolhead{$ d\rm_{helio} $}	&\colhead{$ x$}&\colhead{$ y $} & \colhead{$ z $} &	\colhead{$v_{x} $}&	\colhead{$v_{y} $}&	\colhead{$v_{z} $}\\ 
		\colhead{}&\nocolhead{(pc)} &\colhead{(pc)}&
		\colhead{(pc)}&
		\colhead{(pc)}&
		\colhead{(\kms)}&
		\colhead{(\kms)}&
		\colhead{(\kms)} 
	}
	\decimalcolnumbers
	\startdata
		CMa & $1128\pm15$ & $-808.5$ & $8935.8$ & $-36.1$ & $-0.7$ & $1.5$ & $-8.2$ \\
		A & $1080\pm41$ & $-766.3$ & $8909.8$ & $-50.8$ & $2.5$ & $-0.1$ & $-9.8$ \\
		B & $1087\pm19$ & $-761.3$ & $8923.9$ & $-48.3$ & $4.5$ & $3.3$ & $-10.2$ \\
		C1 & $1118\pm44$ & $-781.7$ & $8948.3$ & $-39.6$ & $-0.5$ & $4.7$ & $-8.2$ \\
		C2 & $1107\pm35$ & $-765.0$ & $8948.7$ & $-35.7$ & $-2.8$ & $3.6$ & $-6.7$ \\
		D & $1151\pm44$ & $-809.2$ & $8968.3$ & $-26.1$ & $-1.0$ & $2.5$ & $-9.3$ \\
		E & $1195\pm47$ & $-832.6$ & $9007.3$ & $-15.4$ & $-0.6$ & $0.5$ & $-6.1$ \\
		F & $1159\pm39$ & $-832.2$ & $8956.0$ & $-9.9$ & $-2.4$ & $0.8$ & $-6.1$ \\
	\enddata
\end{deluxetable*}

\bibliography{mybib}{}
\bibliographystyle{aasjournal}

\end{document}